\documentclass[useAMS,usenatbib]{mn2e}

\title[Molecular gas inflow towards the Seyfert nucleus of NGC\,4051]{Mapping of molecular gas inflow towards the Seyfert nucleus of NGC\,4051 using Gemini NIFS}

\author[Rogemar A. Riffel et al.]{Rogemar A. Riffel$^{1}$\thanks{E-mail:rogemar@ufrgs.br}, Thaisa Storchi-Bergmann$^{1}$, Cl\'audia Winge$^{2}$,
\newauthor Peter J. McGregor$^{3}$, Tracy Beck$^{4}$  and Henrique Schmitt$^{5}$\\
$^{1}$Universidade Federal do Rio Grande do Sul, IF, CP 15051, Porto Alegre 91501-970, RS, Brazil.\\
$^{2}$Gemini Observatory, c/o AURA Inc., Casilla 603, La Serena, Chile.\\
$^{3}$Research School of Astronomy and Astrophysics, Australian National University, Cotter Road, Weston Creek, ACT 2611, Australia.\\
$^{4}$Gemini Observatory, Northern Operations, Hilo, HI, USA.\\
$^{5}$Remote Sensing Division, Naval Research Laboratory, 4555 Overlook Avenue, SW, Washington, DC 20375, USA; \\Interferometric Inc., 13454 Sunrise Valley, Suite 240, Herndon, VA 20171.}

\usepackage{graphicx}

\begin{document}

\date{Accepted 1988 December 15. Received 1988 December 14; in original form 1988 October 11}

\pagerange{\pageref{firstpage}--\pageref{lastpage}} \pubyear{2002}

\maketitle

\label{firstpage}

\begin{abstract}
We present two-dimensional (2D) stellar and gaseous kinematics of the inner $\sim$\,130$\times$180 pc$^2$ of the Narrow Line Seyfert 1 galaxy NGC\,4051 at a sampling of 4.5\,pc, from near-infrared $K$-band spectroscopic  observations obtained with the Gemini's Near-infrared Integral Field Spectrograph (NIFS) operating with the ALTAIR  adaptive optics module. We have used the CO absorption bandheads around 2.3\,$\mu$m to obtain the stellar kinematics which show the turnover of the rotation curve at only $\approx$55\,pc from the nucleus, revealing a  highly concentrated gravitational potential. The stellar velocity dispersion of the bulge is $\approx$\,60\,km\,s$^{-1}$ -- implying on a nuclear black hole mass of $\approx\,10^6$\,M$_\odot$ -- within which patches of lower velocity dispersion suggest the presence of regions of more recent star formation. From measurements of the emission-line profiles we have constructed two-dimensional maps for the flux distributitions, line ratios, radial velocities and gas velocity dispersions for the H$_2$, H\,{\sc i} and [Ca\,{\sc viii}]  emitting gas.
 Each emission line samples a distinct kinematics. The Br$\gamma$ emission-line shows no rotation as well as no blueshifts or redshifts in excess of 30\,km\,s$^{-1}$, and is thus not restricted to the galaxy plane. 
The [Ca\,{\sc viii}] coronal region is compact but resolved, extending over the inner 75\,pc. It shows the highest blueshifts -- of up to $-250$\,km\,s$^{-1}$, and the highest velocity dispersions, interpreted as due to outflows from the active nucleus, supporting an origin close to the nucleus. Subtraction of the stellar velocity field from the gaseous velocity field has allowed us to isolate  non-circular motions observed in the H$_2$ emitting gas.
The most conspicuous kinematic structures are two nuclear spiral arms -- one observed in blueshift in the far side of the galaxy (to the NE), and the other observed in redshift in the near side of the galaxy (to the SW). We interpret these structures as inflows towards the nucleus, a result similar to those of previous studies in which we have found streaming motions along nuclear spirals in ionized gas using optical IFU observations.  We have calculated the mass inflow rate along the nuclear spiral arms, obtaining $\dot{M}_{H_2} \approx 4\times10^{-5}\,{\rm M_\odot\,yr^{-1}}$, a value $\sim$\,100 times smaller than the accretion rate necessary to power the active nucleus. This can be understood as due to the fact that we are only seeing the hot ``skin'' (the H$_2$ emitting gas) of the total mass inflow rate, which is probably dominated by cold molecular gas.
From the H$_2$ emission-line ratios we conclude that X-ray heating can account for the observed emission, but the H$_2\,\lambda2.1218\,\mu$m/Br$\gamma$ line ratio suggests some contribution from shocks in localized regions close to the compact radio jet. 

\end{abstract}


\section{Introduction}

The presence of supermassive black holes (SMBHs) at the centres of all galaxies which have stellar bulges is nowadays widely accepted by the astronomy comunity \citep{gebhardt00,ferrarese00}. According to this scenario the energy emitted by an Active Galactic Nucleus (AGN) is due to the accretion of material onto the SMBH and implies the presence of a gas reservoir close to the AGN. \citet{lopes07} using archival Hubble Space Telescope (HST) optical images for a large sample of  early-type galaxies  with and without AGNs, found that all AGN hosts have circumnuclear gas and dust, while this is observed in only 26\% of a pair-matched sample of inactive galaxies.   A gas reservoir close to the AGN is also supported by the presence of recent star formation  in the circumnuclear region of active galaxies \citep{schmitt99,boisson00,storchi-bergmann00,cid-fernandes01,cid-fernandes05,storchi-bergmann05}. However, the strongest signatures that feeding to the SMBH is occurring include the observation of streaming motions in ionized gas along nuclear spirals towards the nucleus of nearby active galaxies using two-dimensional (2D) optical spectroscopy \citep{fathi06,storchi-bergmann07}.

Streaming motions as feeding signatures to the nuclear region have been previously observed in radio wavelengths. \citet{adler96}, for example, have found streaming motions towards the center along the spiral arms of M\,81 in H\,{\sc i}, while \citet{mundell99} have found similar streaming motions towards the nucleus along the weak bar of NGC\,4151. Closer to the center, most of the gas is in the molecular phase, and CO observations have been used to map the gas kinematics and inflows  \citep[e.g.][]{garcia-burillo03,krips05,boone07}. 

Molecular hydrogen emission lines are also relatively strong in the near-IR $K$-band spectra of active galaxies, and previous studies suggest that its distribution and kinematics is distinct from that observed in the other emission lines, which are usually dominated by outflows \citep[e.g.][]{crenshaw00,das05}. In the Seyfert galaxies NGC\,2110 and Circinus, for example, \citet{storchi-bergmann99} have found broader emission-line profiles for [Fe\,{\sc ii}] and Pa$\beta$ than for the H$_2\lambda$2.1218$\mu$m emission-line in a study using near-IR long-slit observations, suggesting a stronger influence of nuclear outflows on the former emission lines and a different origin for the H$_2$ emitting gas, consistent with the colder kinematics of the galaxy disk. More recently, using two-dimensional (hereafter 2D) near-IR spectroscopy of the Seyfert galaxy ESO\,428-G14, we \citep{riffel06} have found that the H$_2$ emission distribution was mostly restricted to the plane of the galaxy and was less affected by the AGN outflow than the [Fe\,{\sc ii}] and Pa$\beta$ emission lines. 

In this paper we use adaptive optics IFU spectroscopic data obtained with the Gemini's Near-infrared  Integral Field Spectrograph  \citep[NIFS - ][]{mcgregor03}, in the near-IR $K$-band at a sampling of  0$\farcs$1$\times$0$\farcs$1, to investigate the stellar and gaseous kinematics of the inner $\sim 3 \times 4$\,arcsec$^2$ of the nearby active galaxy NGC\,4051. The stellar kinematics will be used to constrain the galaxy potential and mass of the SMBH, but our main goal is to look for signatures of feeding mechanisms at parsec scales through the H$_2$ kinematics.



NGC\,4051 is a SABbc galaxy at a distance of only $\sim$9.3\,Mpc\,\citep{barbosa06},
such that 1$\farcs$0 corresponds to 45\,pc at the galaxy. 
It harbors one of the closest AGN, classified as a Narrow-Line Seyfert 1 (NLS1). 
We have selected NGC\,4051 for this study partially on the basis of the recent work of \citet{barbosa06}, who
obtained 2D stellar kinematics using the IFU of the Gemini Multi-Object 
Spectrograph (GMOS) to observe the stellar absorption lines of the Calcium triplet around 8500\AA.
These authors have  shown that the turnover of the stellar rotation curve occurs at only $R\sim$50\,pc from the nucleus, indicating that the stellar motions are dominated by a highly concentrated gravitational potential.
As NIFS with the ALTAIR adaptive optics module provides a much better image sampling and resolution than the GMOS-IFU, we decided to further investigate the kinematics of the nuclear region of NGC\,4051 in order to
better constrain the gravitational potential and the mass of the SMBH. In adition, this galaxy shows strong H$_2$ emission \citep{rogerio06}, making it a good candidate for a study of its kinematics.


Previous studies of NGC\,4051 include HST narrow-band [O\,{\sc iii}] images 
which show an unresolved nuclear source and faint extended emission (by 1\farcs2)
along the position angle PA=100$^\circ$
\citep{schmitt96}, the approximate direction of the alignment of 
two radio components at 6\,cm separated by 0$\farcs$4
\citep{ulvestad84}. 
\citet{veilleux91} reported that the profiles of optical forbidden emission lines
present blue wings reaching velocities of up to 
800\,km\,s$^{-1}$, and proposed a model for the narrow-line region (hereafter NLR) with outflows and obscuring dust. Evidence for outflows have also been observed by \citet{christopoulou97} for the [O\,{\sc iii}] emission to 1\farcs5 NE of the nucleus.
\citet{nagao00} using the
[Fe\,{\sc X}]$\lambda$6374 emission line report a
high ionization region extending to 3$\farcs$0\ SE of the
nucleus.  \citet{lawrence85} found a X-ray variability on time scales of the order of
1 hour and \citet{salvati93} reported flux changes by a factor of 2 in 6 months
observed at 2.2\,$\mu$m and suggest that emission by dust reprocessing of the UV radiation
from the nucleus is an acceptable explanation.
\citet{ponti06} modelled the
nuclear emission by two power law components, one due to the AGN and other due to reflection by the accretion disk.


This paper is organized as follows: in section 2 we describe the observations
and data reduction. In section 3 we present the results for the stellar kinematics. In section 4 we present the emission-line flux distributions and in section 5 we present the results for the gas kinematics. In section 6 we discuss the results and in section 7 we present the conclusions of this work.

\section{Observations and Data Reduction}

\begin{figure}
\centering
\includegraphics[scale=0.58]{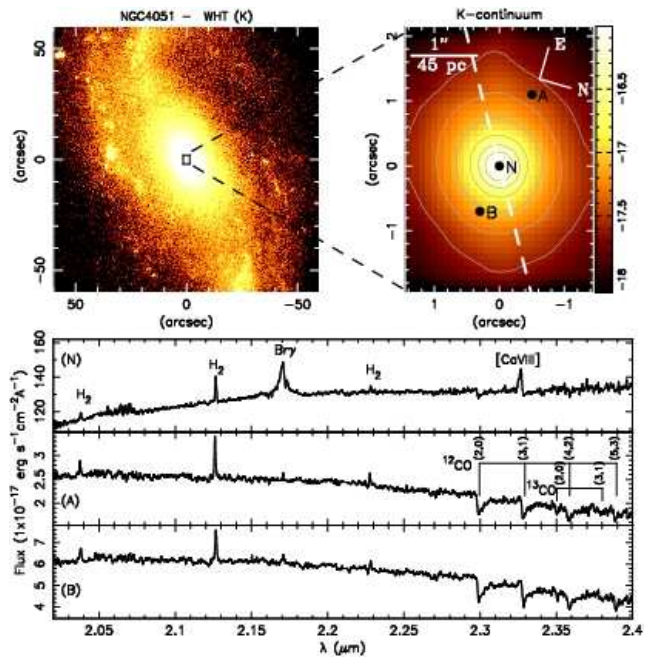}
\caption{Top left: WHT K-band large scale image of NGC\,4051 from \citet{knapen03}. The image has been rotated to the same orientation of the NIFS observations. The box represents the IFU field of view. Top right: $2.12\,\mu$m continuum map from IFU spectroscopy. Bottom: Spectra at the positons N, A and B marked at the top-right panel with the emission lines and CO absorption bandheads identified.}
\label{espectro}
\end{figure}

The IFU spectroscopic data were obtained with NIFS 
operating with the ALTAIR adaptive optics module on the 8-m Gemini North telescope in January 2006 under the instrument
science verification program GN-2006A-SV-123. The IFU has a square field of
view of $\approx3\farcs0\times3\farcs0$, divided into 29 slices with an
angular sampling of 0$\farcs$1$\times$0$\farcs$04. The observing procedures
followed the standard Object-Sky-Sky-Object dither sequence, with off-source
sky positions since the target is extended, and individual exposure times of
750\,s centered at $\lambda = 2.2499\,\mu$m. Two set of observations, each with three individual exposures, were
obtained at different spatial positions; the first one centered at the
position 0$\farcs$4  from the nucleus along the PA$=-74^\circ$ and the second at a position offset 0$\farcs$5 along PA$=106^\circ$. The longest extent covered by the IFU 
observations was oriented along the position 
angle PA$=106^\circ$, which corresponds to the orientation of the line of
nodes derived by modelling of the stellar velocity field with a rotating disk
\citep{barbosa06}. 
We have used the $\rm K_-G5605$ 
grating and the filter $\rm  HK_-G0603$, which resulted in an arc lamp line full width at half maximuum (FWHM) of 
 3.2\,\AA.

The data reduction was accomplished using tasks contained in the {\sc nifs}
package which is part of {\sc gemini iraf} package as well as generic {\sc
iraf} tasks. The reduction procedure included trimming of the images,
flat-fielding, sky subtraction, wavelength and s-distortion calibrations. We
have also removed the telluric bands and flux calibrated the frames by
interpolating a black body function to the spectrum of the telluric standard star.
The final IFU data cube contains 1160 spectra, each spectrum corresponding to an angular coverage of
0$\farcs$1$\times$0$\farcs$1, which translates into 4.5$\times$4.5\,pc$^2$ at the galaxy and
covering the spectral region from 1.991$\,\mu$m to 2.425$\,\mu$m. The total observed
field of view 2$\farcs$9$\times$4$\farcs$0 (obtained by mosaicing the two set of observations) thus corresponds to a region of projected dimensions 130\,pc$\times$180\,pc at the galaxy.  

In the top-left panel of Fig.\,\ref{espectro} we present a large scale $K$-band image from \citet{knapen03} obtained at the William Hershel Telescope (WHT),
The central rectangle shows the IFU field of view. The large scale image was rotated to the same orientation of the IFU observations.  In the top-right panel we present an image obtained from the NIFS data cube for the  2.12$\,\mu$m continuum emission, obtained by interpolation of the continuum under the H$_2\,\lambda2.1218\,\mu$m emission line. In the bottom panels we present three charateristic IFU spectra: the nuclear spectrum (position N in the continuum map), a spectrum from a location at 1\farcs2 E of the nucleus (position A) and another from 0\farcs75 W of the nucleus  (position B). Both spectra correspond to an aperture of  0$\farcs$3$\times$0$\farcs$3. The emission lines are identified in the nuclear spectrum: the H$_2$ lines at $\lambda=$2.0338, 2.1218 and 2.2235\,$\mu$m, the H\,{\sc i} Br$\gamma$ at 2.1661\,$\mu$m  and the [Ca\,{\sc viii}] coronal emission line at 2.3211\,$\mu$m. The $^{12}$CO and $^{13}$CO absorption bandheads used to obtain the stellar kinematics are identified in the spectrum from position A.

\section{Line of sight velocity distribution}


In order to obtain the line-of-sight velocity distribution (LOSVD) we have
used the penalized Pixel-Fitting (pPXF) method of
\citet{cappellari04} to fit the stellar 
absorptions  present in the $K$-band
spectra. The algorithm finds the best fit to a galaxy spectrum by convolving a
template stellar spectrum with the corresponding LOSVD. This procedure gives
as output the radial velocity, velocity dispersion  and
higher-order Gauss-Hermite moments. The pPXF method allows the use of several template stellar spectra and to vary the weights of the contribution of the different templates to obtain the best fit, minimizing the template mismatch problem. However, the use of pPXF
requires templates which match closely the galaxy spectrum. 
\citet{emsellem04} present a extensive discussion about the use of the pPXF method and the templates
mismatch problem and \citet{silge03} present a discussion about template mismatch for kinematic fitting using the CO absorption bandheads in the near-IR. 

A high signal-to-noise ratio  is required for reliable stellar kinematic measurements using the pPXF method. In previous works the data are usually binned to give S/N ratio between 40 and 60 over the whole field-of-view. In our data, ratios smaller than these are observed only very close to the borders of the field (S/N $\approx$ 35 measured blueward of the first CO band-head). Our typical S/N ratios  are 80 with maximuum values reaching 120 close to the nucleus, thus allowing  reliable kinematic measurements without spatial binning of the data.


\subsection{The stellar templates}

In this study we have selected as template spectra, those of the spectroscopic library of late spectral type stars observed with the  Gemini Near Infrared Spectrograph (GNIRS) IFU using the grating 111 l/mm in the {\it K}-band \citep{winge07}. This library is composed of spectra of 29 objects, which include dwarf, giant and sub-giant stars with spectral types from F7\,{\sc iii} to M3\,{\sc iii}, observed in the spectral range from 2.24 to 2.42\,$\mu$m. Of these, 23 stars were also observed on a second setting extending the wavelength coverage down to 2.15\,$\mu$m. 
The spectra of the templates have signal-to-noise ratios larger than 50 blueward of the CO\,2-0 first-overtone bandhead.  Detailed discussion about this library is presented by \citet{winge07}. The spectral resolution of the stellar spectra is 3.3\,\AA~ -- very close to the spectral resolution of the NIFS data, thus the library can be used to fit the stellar kinematics without any resolution correction. We have verified this in some tests using available NIFS spectra of a few stars. We opted to use the GNIRS library because there are too few available template spectra with NIFS.
We have chosen 16 stars from the library which better match the stellar features of NGC\,4051.



In order to investigate the influence of the stellar templates on the 
velocity dispersions  obtained we have fitted the stellar kinematics using individual stars as
templates.  We observed that the large scale structures in $\sigma$ maps are similar for all stars, but the mean $\sigma$ values vary significantly -- higher EWs in the templates result in lower $\sigma$ values for the galaxy.
This result evidences the  importance of using a large library of stellar templates in order to obtain reliable velocity dispersion measurements.



\subsection{The effect of the [Ca\,{\sc viii}] coronal line} 


The nuclear spectrum of NGC\,4051 (Fig.\,\ref{espectro}) shows the coronal
emission line of [Ca\,{\sc viii}] at 2.3211$\,\mu$m. \citet{davies06} have shown that this
line affects the kinematic measurements obtained from the
CO bandheads. In order to investigate the influence of this line in our
measurements, we have chosen two spectral regions to fit the stellar
kinematics: the first one (2.258-2.372\,$\mu$m) includes the [Ca\,{\sc viii}] line and the
second (2.258-2.314\,$\mu$m and 2.346-2.372\,$\mu$m) excludes this line (together with the ``contaminated'' CO bandhead). We observed that the radial velocities and velocity dispersions derived using the first spectral range are higher in the nuclear region, where the [Ca\,{\sc viii}] emission line is present,  than those obtained by using the second spectral region. We thus decided to exclude the contaminated $^{12}$CO\,3-1 bandhead  from the stellar kinematics fitting in the region where the [Ca\,{\sc viii}] emission line is present (the central 0\farcs8 radius). In regions away from the nucleus we used the first spectral range which includes the $^{12}$CO\,3-1 bandhead.


\subsection{The stellar kinematics}

\begin{figure*}
\centering
\begin{minipage}{1\linewidth}
\includegraphics[scale=0.5]{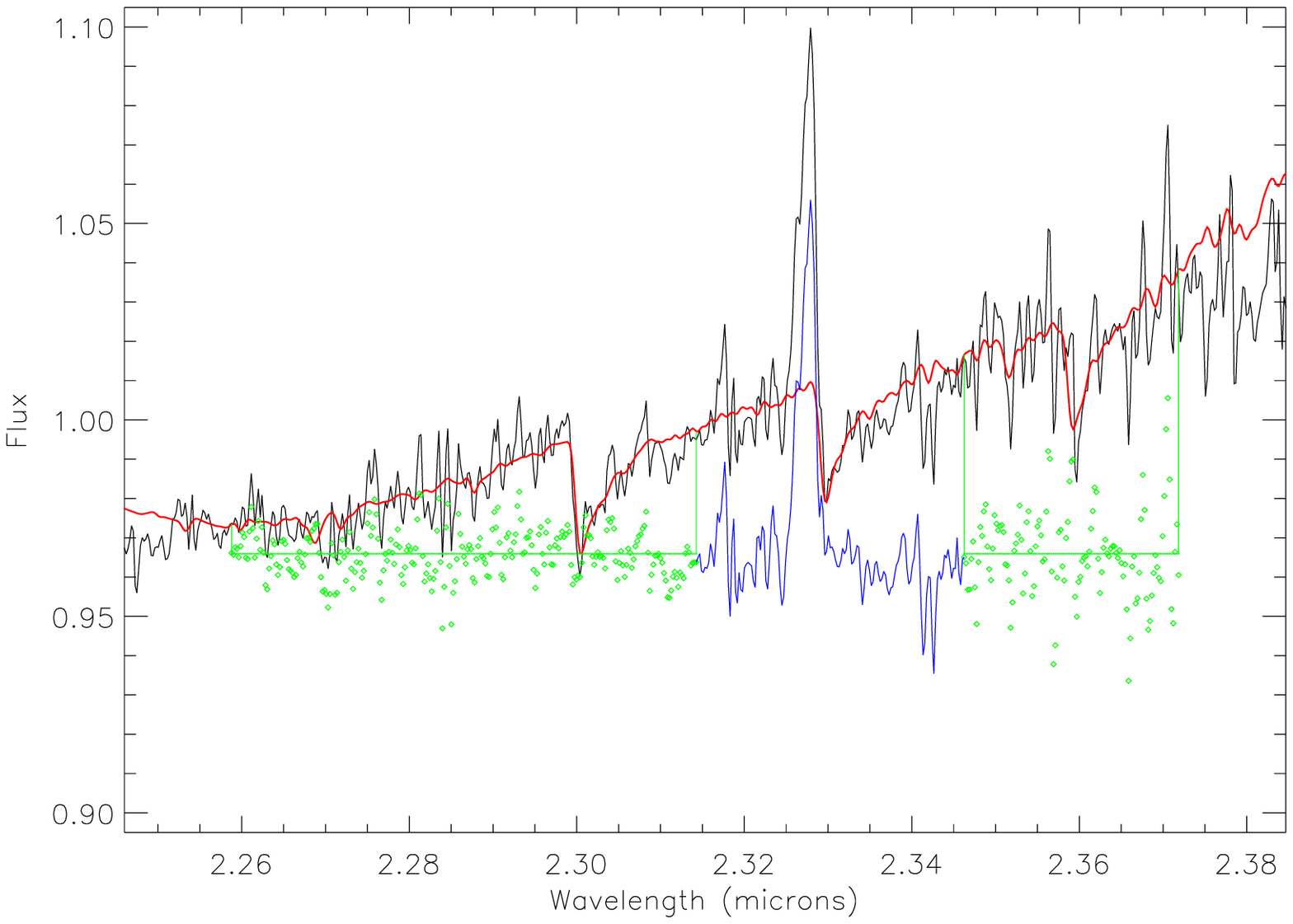}
\includegraphics[scale=0.5]{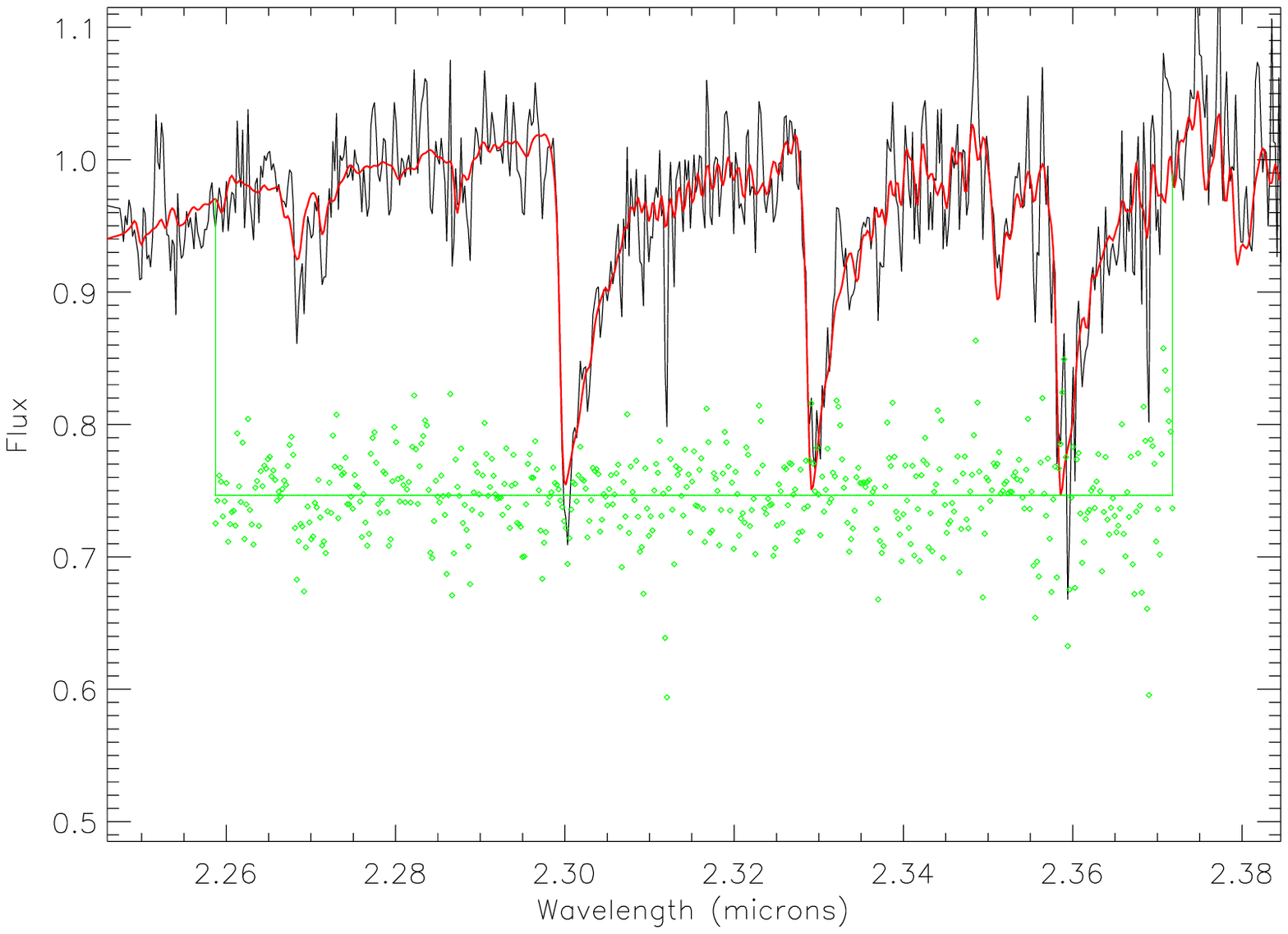}
\end{minipage}
\begin{minipage}{1\linewidth}
\includegraphics[scale=0.5]{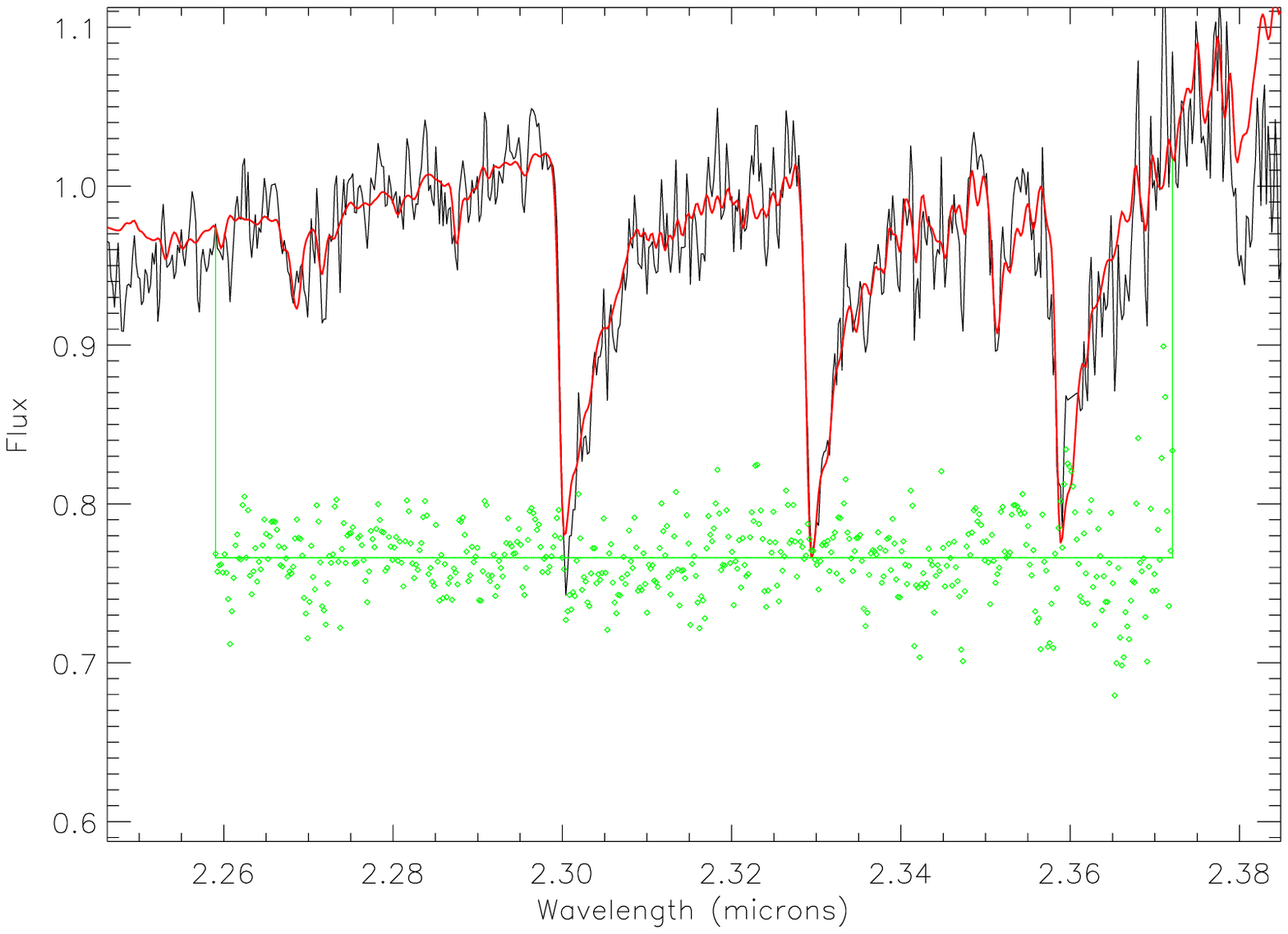}
\includegraphics[scale=0.5]{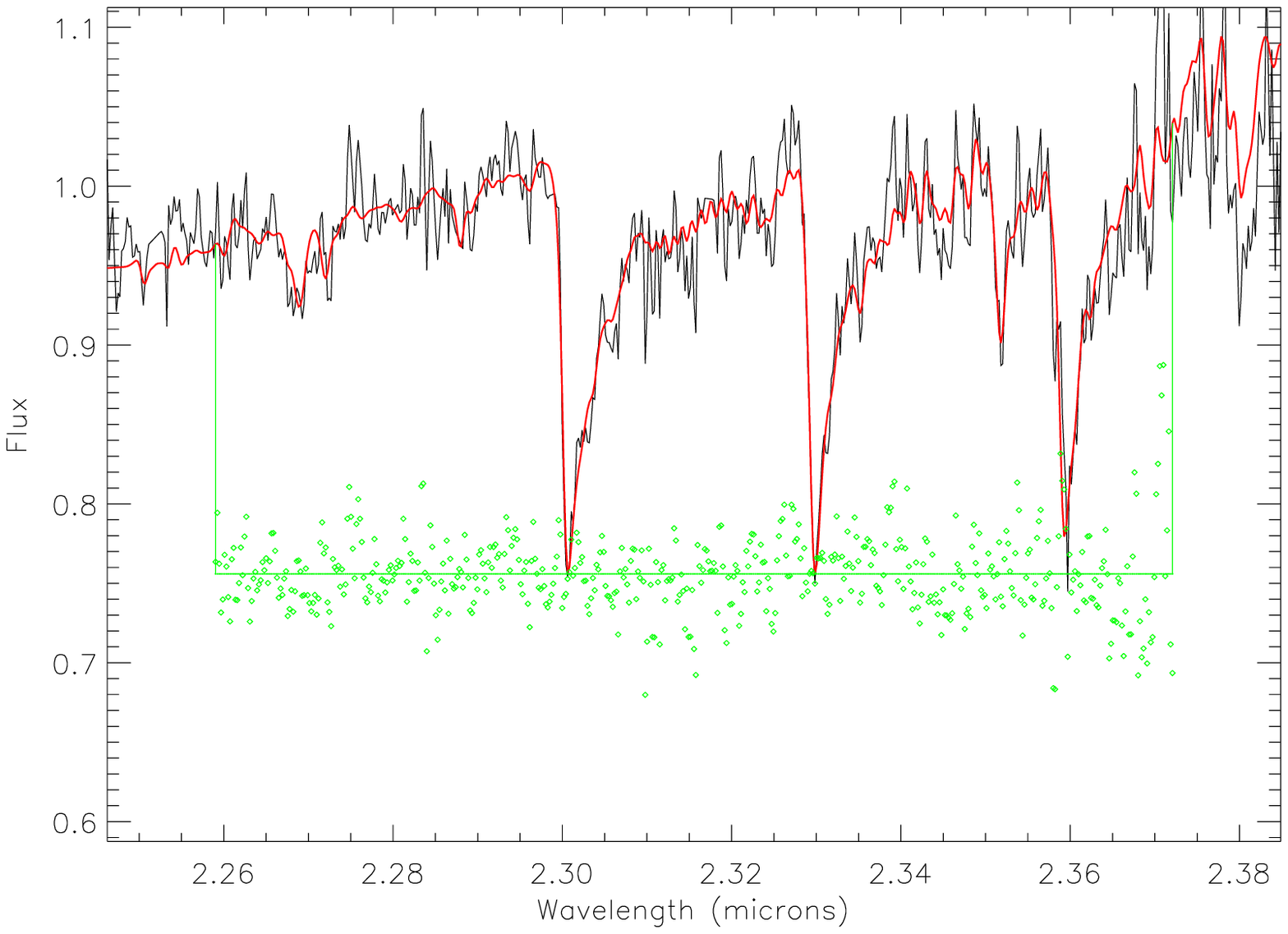}
\end{minipage}
\caption{Sample fits of the stellar kinematics of the nuclear region of NGC\,4051 using pPXF. Top left: fit of the nuclear spectrum; top right: fit of the spectrum at 1$\farcs$5 E of the nucleus; bottom left: fit of the spectrum at the location corresponding to the blue turnover of the rotation curve at 1$\farcs$2 SE of the nucleus; bottom-right: fit of the spectrum at the position of the red turnover of the rotation curve at 1$\farcs$2 NW of the nucleus.  The observed spectra are shown in black, the fits in red, the residuals in green, while in blue we show the spectral range not included in the fit of the central region.}
\label{fit}  
\end{figure*}

In Fig.\,\ref{fit} we show the fits of the stellar templates to the galaxy
spectra with the program pPXF for four different positions: the nucleus; 1$\farcs$2 SE and 1$\farcs$2\,NW of the nucleus, where we observe the turnovers of the rotation curve and a location at 1$\farcs$5 E of the nucleus, almost at the border of the NIFS field. We observe that the stellar templates fit very well the galaxy spectra at most positions, including  regions near the border of the IFU field, where the signal-to-noise ratio is smaller. 
The results for the nucleus should nevertheless be considered with caution, as the fitting may have been affected by dust emission and emission lines present in the galaxy spectrum but absent in the stellar templates.



\begin{figure*}
\centering
\includegraphics[scale=1.4]{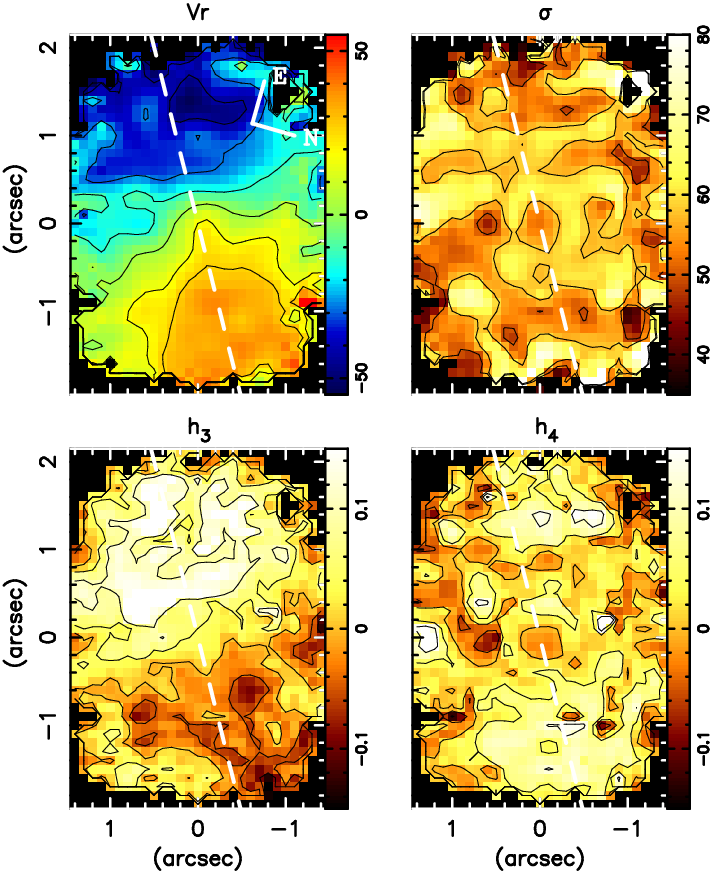}
\caption{Stellar kinematic maps obtained from the pPXF fit. Top: radial velocity (left) and velocity dispersion (right) maps. Bottom: $h_3$ and $h_4$  Gauss-Hermite moments. The mean uncertainties are 10 km\,s$^{-1}$ for radial velocity,  8 km\,s$^{-1}$ for $\sigma$, and 0.03 for $h_3$ and $h_4$. The dashed lines show the position of the line of nodes.}  
\label{stel}
\end{figure*}

\begin{figure}
\centering
\includegraphics[scale=0.45]{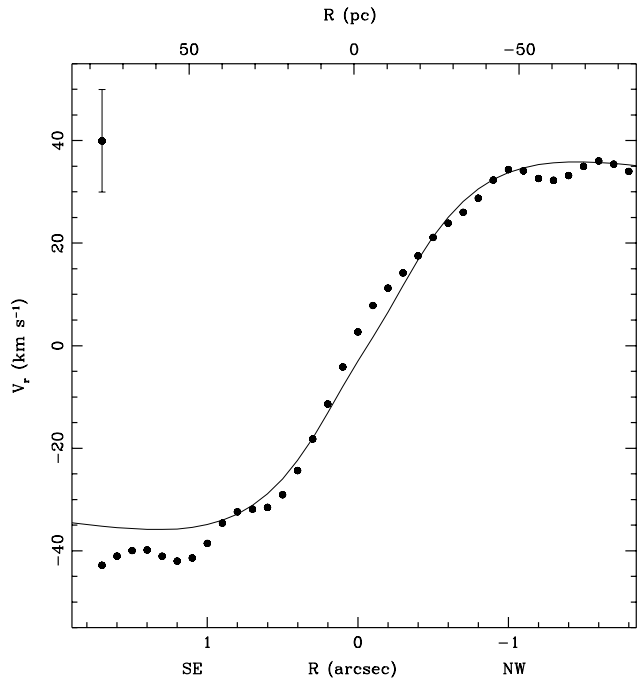}
\caption{One dimensional stellar rotation curve for a pseudo-slit oriented along the major axis of the galaxy. The points are the observed radial velocities and the full line is the modeled velocities.}  
\label{corte}
\end{figure}

In Fig.\,\ref{stel} we present the resulting stellar
kinematics. The black regions in this figure are masker regions where the
signal to noise ratio in the spectra was too low to provide a reliable fit. In the top-left panel we
show the stellar velocity field, which shows a velocity range from $\approx -40$ to $\approx 40$\,km\,s$^{-1}$ with the turnover occurring at $\approx$55\,pc from the nucleus in good agreement with the velocity field obtained by \citet{barbosa06} from optical IFU data. The mean uncertainties in the velocities are
$\approx{\rm 10\,km\,s^{-1}}$. The turnover and amplitude of the rotation curve can be more easily observed in 
the one dimensional cut of the stellar velocity field shown in Fig.\,\ref{corte}. 
The one dimensional rotation curve was obtained by the averare of the velocities within a pseudo-slit with 1\,arcsec width oriented along the major axis of the galaxy.

The stellar velocity dispersion map is shown in the top-right panel of Fig.\,\ref{stel}. The $\sigma$ map presents values ranging from
$\approx$35 to $\approx$90\,km\,s$^{-1}$, with mean uncertainties of
$\approx{\rm 8\,km\,s^{-1}}$.  The bottom panels  show the higher order Gauss-Hermite moments $h_3$ (left) and $h_4$ (right).   These moments measure devations of the line profile from  a  Gaussian: the parameter $h_3$ measures asymmetric deviations and the $h_4$ measure symmetric deviations \citep{marel93}. The values $h_3$ and $h_4$ vary from -0.15 to 0.15 with mean uncertainties of 0.03.  The highest values of $h_3$ are observed to SE of the nucleus and the lowest values to NW of the nucleus. The $h_4$ values are nearly zero over most of the IFU field.

\subsection{Kinematic Modelling}\label{secplum}

Since the stellar velocity field is dominated by rotation, it was fitted with a velocity model produced by a Plummer potential in order to obtain  the systemic velocity, orientation of the line of nodes, bulge mass and the position of the kinematical center. The Plummer potential is given by:
\begin{equation}
\Phi=-\frac{GM}{\sqrt{r^2+a^2}},
\end{equation}
where  $a$ is a scale length, $r$ is the radial distance in the plane of the galaxy, $M$ is the mass inside $r$ and $G$ is the Newton's gravitational constant. Defining the coordinates of the kinematical center of the system as ($X_0,Y_0$), the observed radial velocity at  position ($R,\Psi$), where $R$ is the
projected radial distance from the nucleus in the plane of the sky and
$\Psi$ is the corresponding position angle, is  given by \citep{barbosa06}:

\begin{equation}
V_r=V_s + \sqrt{\frac{R^2GM}{(R^2+A^2)^{3/2}}}\frac{{\rm sin}(i){\rm cos}(\Psi-\Psi_0)}{\left({\rm cos^2}(\Psi - \Psi_0) + \frac{{\rm sin^2}(\Psi-\Psi_0)}{{\rm cos^2}(i)} \right)^{3/4}}
\end{equation}
where $V_s$ is the systemic velocity, $i$ is the inclination of the
disk ($i=0$ for face on disk) and $\Psi_0$ is the position angle of
the line of nodes. The relations between $r$ and $R$, and between $a$ and $A$ are: $r=\alpha R$ and $a=\alpha A$, where $\alpha=\sqrt{{\rm{cos^2}(\Psi-\Psi_0)+\frac{{\rm sin^2}(\Psi-\Psi_0)}{{\rm cos}(i)}}}$. The equation above
contains six free parameters, including the kinematical center, which
can be determined by fitting the model to the observations. This was
done using a Levenberg-Marquardt least-squares fitting algorithm, in which initial
guesses are given for the free parameters. As the inclination of the disk is tightly coupled with $M$ as $V_r^2 \propto M {\rm sin}(i)$, it cannot be left as a free parameter. We have adopted the value of $i=41.4^\circ$,  
an estimate obtained from ${\rm cos}(i)=\frac{b}{a}$, where $b$ and $a$ are the semi-minor and semi-major axis of the large scale disk as quoted in the  NASA/IPAC Extragalactic Database (NED) for this galaxy.

The parameters derived from the fit are: the systemic velocity corrected by the observatory motion relative to the local standard of rest $V_s=$716\,$\pm$\,11\,km\,s$^{-1}$, $\Psi_0=$120$^\circ$\,$\pm$\,1$^\circ$, $M=$7.7$\pm$0.6\,$\times$\,10$^7$\,M$_\odot$ and $A=$39.7\,$\pm$\,2.7\,pc. The derived kinematical center is very close to the peak of the continuum emission, with $X_0$=4.9$\pm$1.4\,pc and $Y_0$=3.1$\pm$1.2\,pc, where $X_0$ and $Y_0$ are measured in relation to the location corresponding to the peak of the continuum. 

\begin{figure*}
\centering
\includegraphics[scale=1.4]{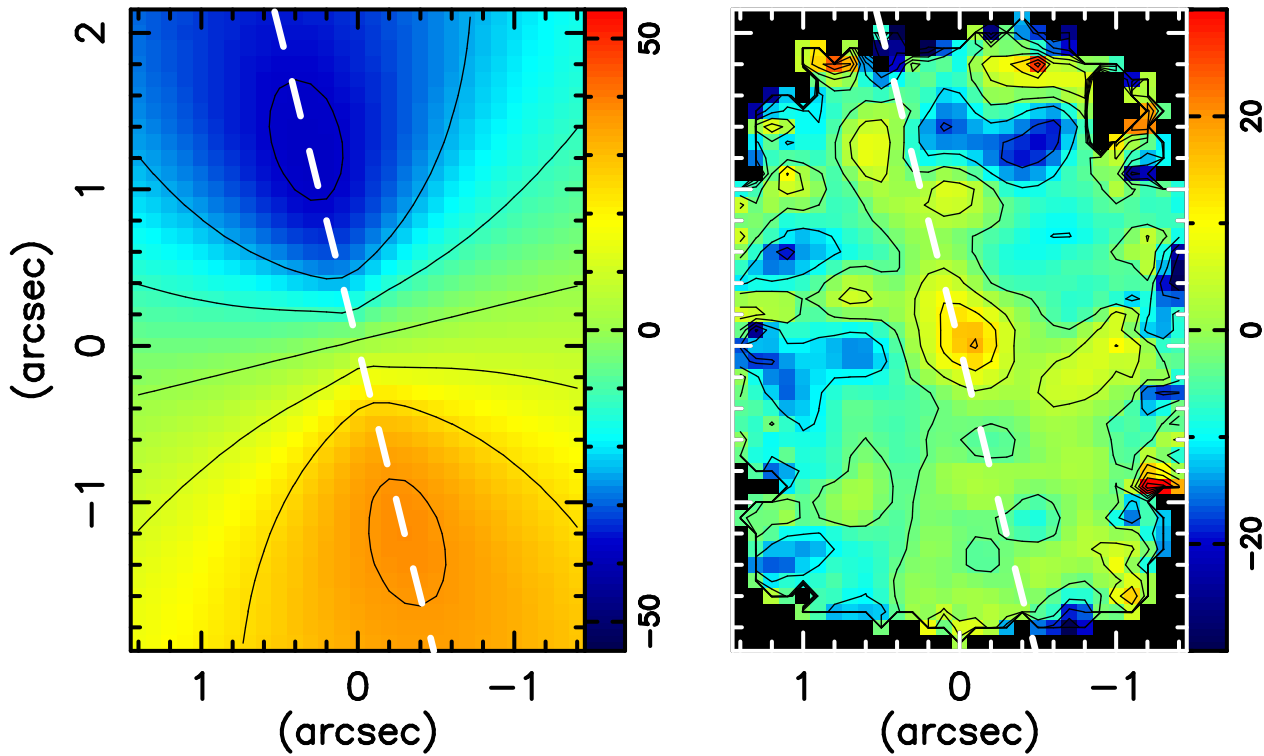}
\caption{Rotating disk model for the stellar kinematics of NGC\,4051 (left) and residual map -- observed velocity field less model (right).  The dashed lines marks the position of the line of nodes.} 
\label{plum}

\end{figure*}

In Fig.\,\ref{plum} we present the derived rotation model in the left panel and the residuals of the stellar velocity field (observed minus modelled) in the right panel. We conclude that the stellar velocity field is well described by the Plummer potential -- the residuals are close to zero over most of the IFU field. The highest residuals ($\sim20$\,km\,s$^{-1}$) are observed within $\sim$0\farcs3 from the nucleus, where the stellar kinematics fitting may have been affected by emission by dust and emission lines.


\section{Emission-line flux distributions}

\begin{figure*}
\centering
\includegraphics[scale=1.15]{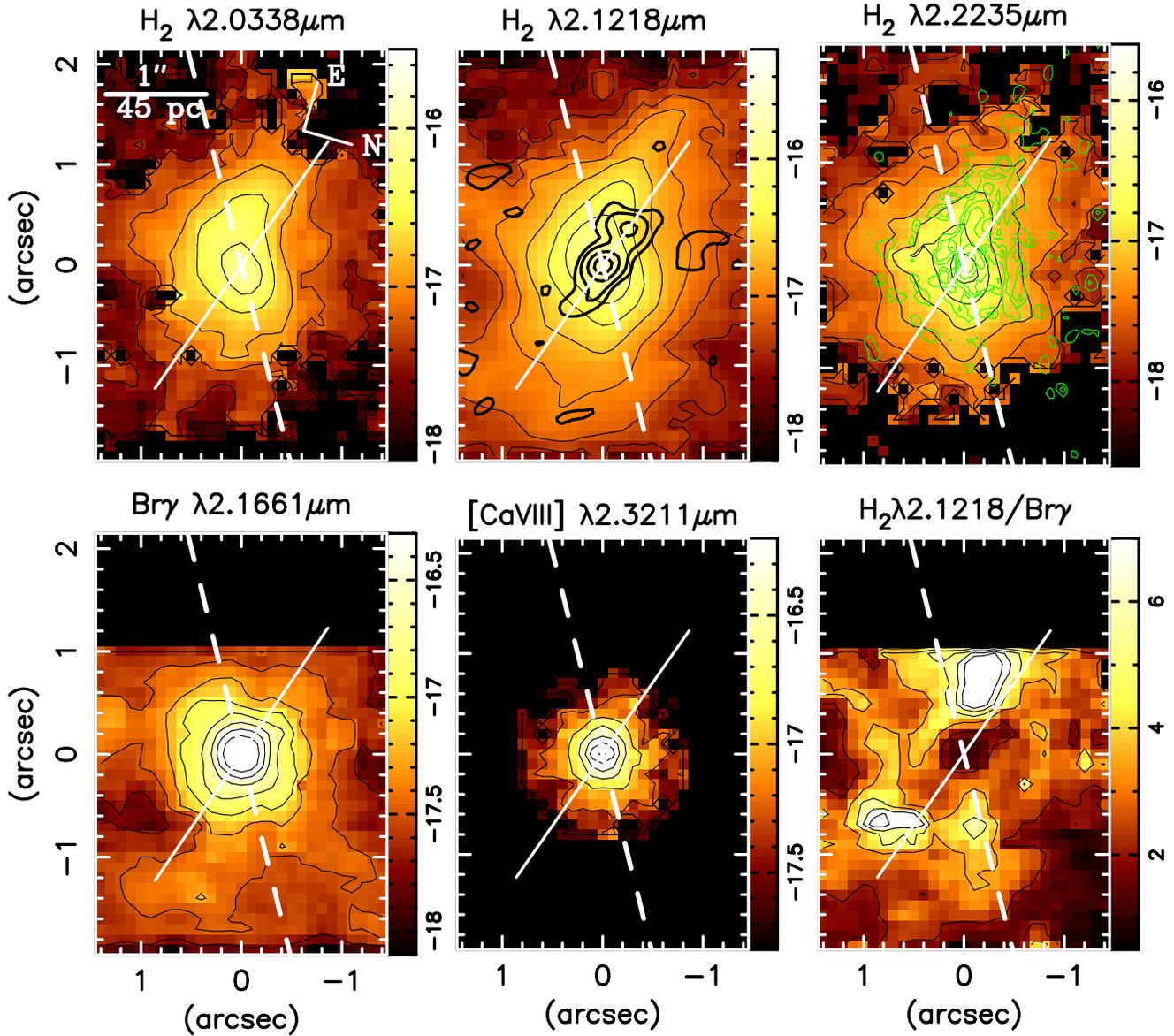}
\caption{Top) From left to right: H$_2\,\lambda$=2.0335, 2.1218 and 2.2235\,$\mu$m  flux maps;  Bottom) left: Br$\gamma$ flux distributions; middle: [Ca\,{\sc viii}]$\lambda$2.3211$\,\mu$m flux distributions and right: H$_2\,\lambda$2.1218/Br$\gamma$ line ratio. The thick black contours overlaid to the H$_2\,\lambda$2.1218$\,\mu$m intensity map are from the VLA radio 3.6\,cm continuum image, thin black lines are isointensities contours for each panel and the green contours are from an [O\,{\sc iii}] narrow-band image from HST. The spatial scale and orientation shown at the top-left panel are the same for all panels. The dashed white line represent the line of nodes of the stellar velocity field and the full white line represent the PA which connects the two radio emission peaks.}
\label{flux}
\end{figure*}

We have fitted Gaussians to the observed emission-line profiles in order to obtain the integrated flux, radial velocity (from the central wavelength of the line) and velocity dispersion (from
the width of the line). 
The corresponding flux maps are shown in Fig.\,\ref{flux}.
 The H$_2$ $\lambda=$2.0338, 2.1218 and 2.2235\,$\mu$m flux distributions are presented in the top left, middle and right panels, respectively, with mean uncertainties of 16\%, for the first one, 5\% for the second one and 9\% for the third. Black regions  identify locations where the line fitting failed due to low signal to noise ratios. The molecular hydrogen emission is extended over most of the observed field. The highest flux values are observed at the nucleus, defined as the location of the peak of the continuum emission. The H$_2$ distribution is extended towards the NE, between the direction of the radio axis (adopted as the one which connects the two radio peaks) and the line of nodes.
The H$_2$ flux distribution shows also a good agreement with the [O\,{\sc iii}] narrow band image from \citet{schmitt96}, whose contours are overplotted in green on the top right panel of Fig.\,\ref{flux}. 

The flux map in the Br$\gamma$ narrow component is shown in the bottom left panel of Fig.\,\ref{flux}. In the central region ($r<25$\,pc) we have fitted two Gaussian components in order to separate the narrow and broad emission line contributions, while in regions further away from the nucleus a single Gaussian was enough for the fit. The signal-to-noise ratio in the top $\sim1^{\prime\prime}$ of the observed field was not high enough to allow the measurement of this line and this is why the region appears black in Fig.\,\ref{flux}.  The mean flux uncertainty for Br$\gamma$ is 24\%. The Br$\gamma$ flux distribution  has the highest flux values at the nucleus and is approximately symmetric around the nucleus, not showing any elongation as observed in the H$_2$ emitting gas.

The flux map of the [Ca\,{\sc viii}]\,$\lambda2.3211\,\mu$m coronal emission line is presented in the bottom middle panel, where the mean flux uncertainties are 15\%. The [Ca\,{\sc viii}] emission peaks at the nucleus and is resolved, extending up to  0\farcs8 from the nucleus, which corresponds to a projected distance of 36\,pc at the galaxy.


The line ratio H$_2\,\lambda$2.1218$\,\mu$m/Br$\gamma$ (only narrao component for Br$\gamma$) can be used as a diagnostic for the excitation mechanism of the molecular hydrogen emission lines \citep[eg.][]{riffel06}, with higher H$_2$/Br$\gamma$ ratios being interpreted as a larger contribution from shocks or X-ray to the H$_2$ excitation. We present this line ratio map in the bottom-right panel of Fig.\,\ref{flux}. The lowest values are      H$_2\,\lambda$2.1218$\,\mu$m/Br$\gamma$ $\approx$1 observed at the nucleus and to the N. The highest values reach H$_2\,\lambda$2.1218$\,\mu$m/Br$\gamma\approx$8 and are observed predominantly in two regions, one aproximately 1\farcs0 W of the nucleus and another at 0\farcs8 E of the nucleus. We note that these two regions are close to the tips of the compact 3.6\,cm radio structure (black contours in the top middle panel). Nevertheless, such a direct interpretation of this map should be considered with caution, as the flux distributions and kinematics (see next section) are obviously different for the H$_2$ and Br$\gamma$ emission lines, implying that may originate in different regions of the galaxy. 

\section{Gas kinematics}

\begin{figure*}
\centering
\includegraphics[scale=1.4]{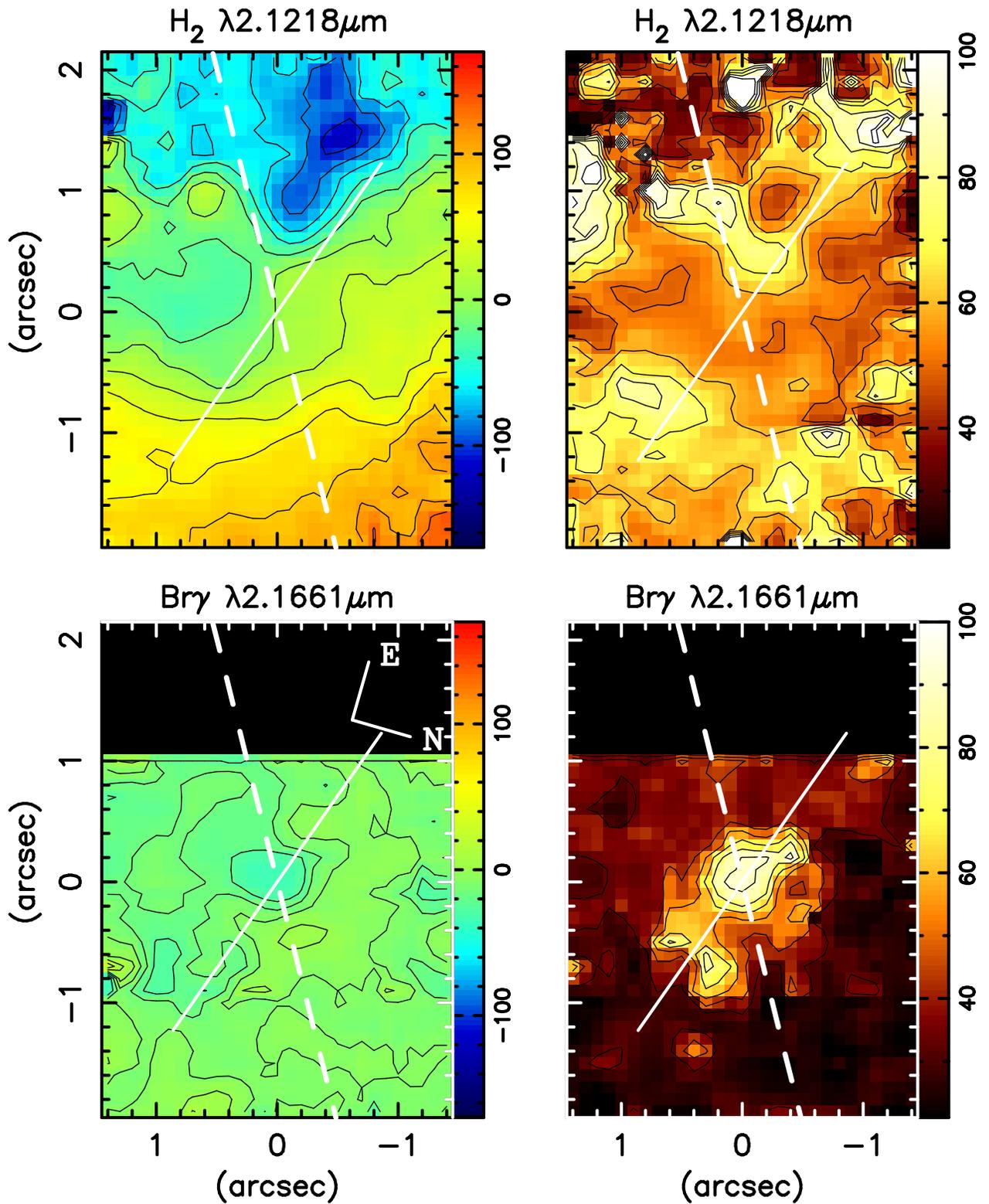}
\caption{Left: Velocity field for the H$_2\,\lambda$2.1218$\,\mu$m (top) and Br$\gamma$ (bottom) emission lines. Right: Velocity dispersion maps for the same emission lines. The contours and lines are as described in Fig.\,\ref{flux}, as well as the spatial scale.}
\label{gas}
\end{figure*}

In the left panels of Fig.\,\ref{gas} we present the radial velocity field obtained from the central wavelengths of the H$_2\,\lambda$2.1218$\,\mu$m and Br$\gamma$ emission lines, with mean uncertainties of 4\,km\,s$^{-1}$ and 9\,km\,s$^{-1}$, respectively. We chose the H$_2\,\lambda$2.1218$\,\mu$m line to represent the  H$_2$ velocity as it is stronger and thus present smaller uncertainties in the measurements than the other H$_2$ emission lines. The systemic velocity of the galaxy, derived from the stellar kinematics modelling, has been subtracted from all the emission line velocity plots.

The H$_2$ velocity field shows a ``rotation pattern'' similar to that of the stars, with the NW side receeding and the SE side  approaching,  although it is quite clear that there are other important kinematic components, evidenced by large deviations from simple rotation. Particularly conspicuous is a blueshifted region to the NE, showing velocity values up to $\approx{\rm -100\,km\,s^{-1}}$, extending by $\sim$1$^{\prime\prime}$\, from the nucleus. 
The Br$\gamma$ velocity field shows no rotation and the total velocity range is only $\approx$50\,km\,s$^{-1}$. 

 In the right panels of Fig.\,\ref{gas} we present the velocity dispersion ($\sigma$) maps obtained from measurements of the FWHM of the emission lines, such that $\sigma=\frac{\rm FWHM}{2.355}$. The $\sigma$ values were corrected for the instrumental broadening and the mean uncertainties are 6\% for H$_2$ and 22\% for Br$\gamma$. The H$_2\,\sigma$ map has values in the range   $\sim$40--100\,km\,s$^{-1}$. A partial ring of low velocity dispersion values ($\sigma \approx 45\,{\rm km s^{-1}}$) is observed surrounding the nucleus, while higher values ($80\leq\sigma\leq100\,{\rm km\,s^{-1}}$) are observed over most of the remaining field. The Br$\gamma$ $\sigma$ map presents the highest values of up to $\approx{\rm 100\,km\,s^{-1}}$ at the nucleus -- but there are uncertainties in these $\sigma$ values due to the broad component contribution -- and at $\approx$0.8\,arcsec to the SW, close to the western tip of the 3.6\,cm radio structure. At positions away from the nucleus the Br$\gamma$ $\sigma$ is lower than 40\,km\,s$^{-1}$.

\subsection{Gas ``Tomography''}
\begin{figure*}
\centering
\includegraphics[scale=1.45]{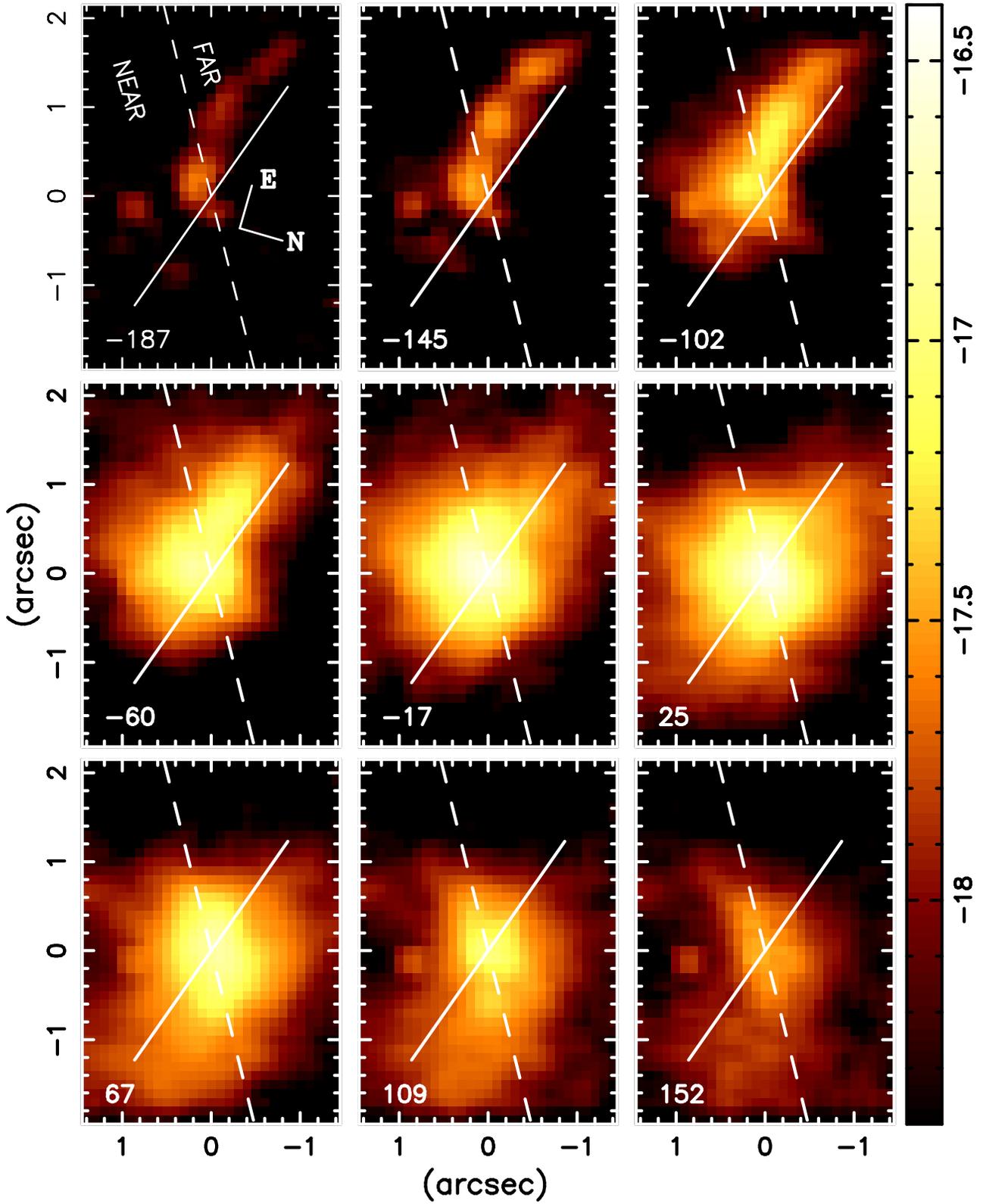}
\caption{Velocity slices along the H$_2$ profile
  with a velocity bin of $\approx{\rm 42\,km\,s^{-1}}$. The cross marks the
  position of the nucleus and the intensity scale is logarithmic. The zero velocity is the systemic velocity obtained from the stellar kinematics modelling. The near and far side of the galaxy are indicated in the top left panel (under the assumption that the spiral arms are trailing).}
\label{slices}
\end{figure*}

\begin{figure*}
\centering
\includegraphics[scale=1.45]{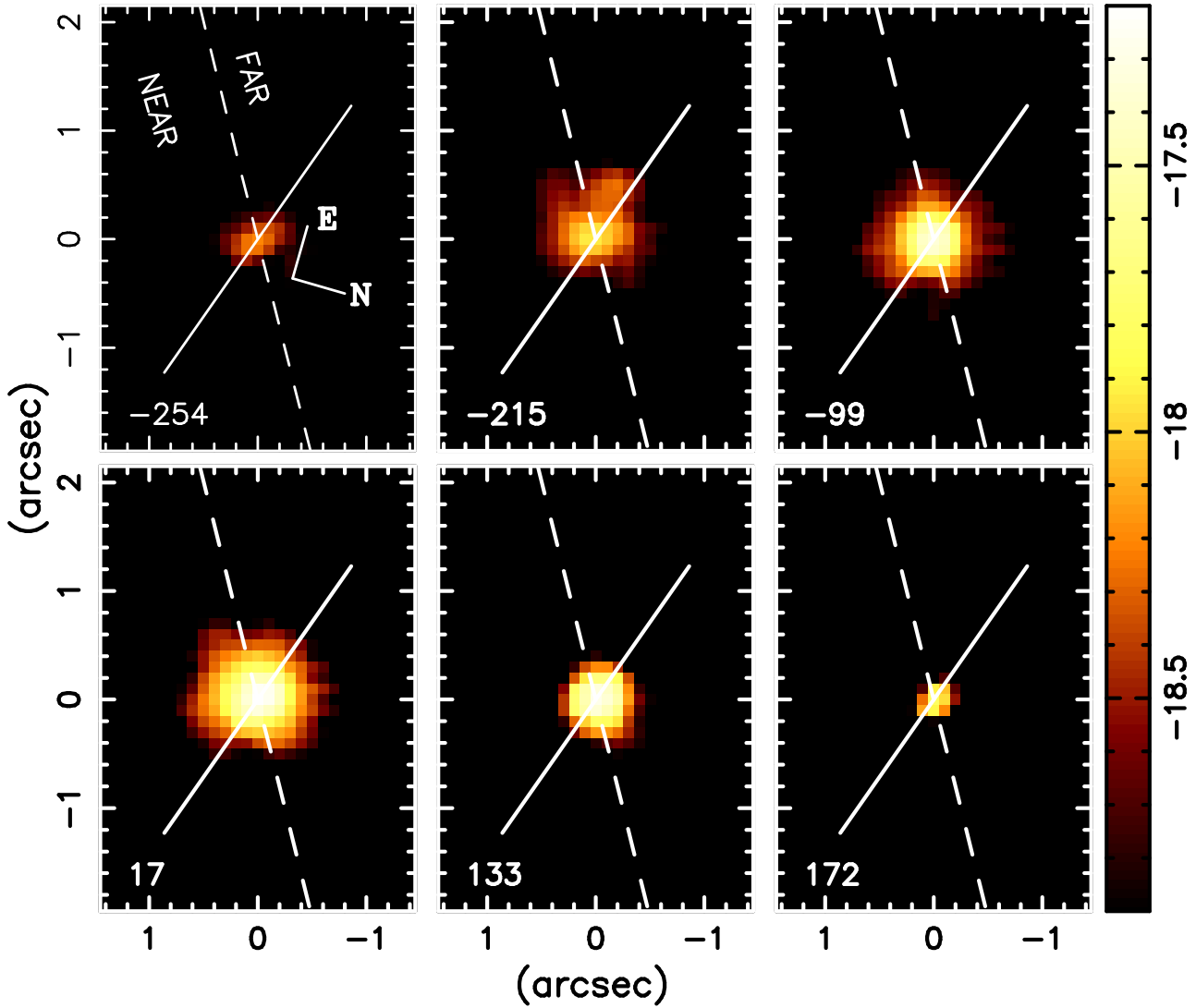}
\caption{Same as Fig.\,\ref{slices} for the [Ca\,{\sc viii}] emission-line profile.}
\label{slices_ca}
\end{figure*} 

The high spectral resolution of the data has allowed us to slice the 
emission-line profiles of H$_2\,\lambda$2.1218$\,\mu$m and  [Ca\,{\sc viii}]\,$\lambda$2.3211$\,\mu$m
into a sequence of velocity bins. With this  ``tomography'' technique,  we can sample the kinematics 
along the whole emission-line profile,
including the wings. In order to obtain the ``tomography" images, we resampled the spectra into bins of 1\,\AA\, with the {\sc scombine} {\sc iraf} task and then combined the two sets of observations into a single data cube using the tasks {\sc scombine} and {\sc imcombine}.  The velocity slices were obtained after  subtraction of the continuum determined as averages of the fluxes from both sides of the emission line. The slices correspond to velocity bins of
$\approx{\rm 42\,km\,s^{-1}}$ (3\,\AA) and are shown in Figs.\,\ref{slices} and \ref{slices_ca}. In these
figures, each panel presents flux levels in logarithmic units for the
velocity slice shown. The zero velocity is adopted as the  systemic velocity of the stars obtained from the stellar kinematic modelling.

The slices trace the gas from negative (blueshifts) to positive (redshifts) velocities relative to the systemic velocity of the stars. For  H$_2$, the highest blueshifts, which reach $\approx{\rm -190\,km\,s^{-1}}$, are observed predominantly to the NE  along a curved elongated structure similar to the one observed in the radial velocity map (top-left panel of
Fig.\,\ref{gas}). This structure -- whose morphology in Fig.\,\ref{slices} can be described as a curved arm -- dominates the emission from $\approx{\rm-190}$ to ${\approx\rm-60\,km\,s^{-1}}$. At velocities close to systemic, the emission is approximately symmetric and dominated by a region of $\approx1^{\prime\prime}$ radius centered at the nucleus. As the slices reach positive velocities, the dominant structure is another curved arm extending to $\approx$2$^{\prime\prime}$ to the W of the nucleus. The highest redshifts reach ${\approx\rm160\,km\,s^{-1}}$.

For [Ca\,{\sc viii}], we show in Fig.\,\ref{slices_ca} the highest velocity bins and exclude a few low velocity bins where the kinematics are similar. The highest blueshifts reach $\approx$-250\,km\,s$^{-1}$, which are higher than those observed for the H$_2$ emitting gas, while the highest redshifts reach velocities of 170\,km\,s$^{-1}$ similar to the ones observed for H$_2$.

We opted not to show slices in the Br$\gamma$ emission-line profile due to the fact that its narrow component is indeed narrow and quite symmetric, and close to the nucleus it is hard to deblend it from the broad line profile, which introduces too much uncertainty in the derived kinematics.


\section{Discussion}

\subsection{Stellar Kinematics} \label{stelkinematics}

As observed in the top-left panel of Fig.\,\ref{stel} {\bf and in Fig.\,\ref{corte}} the turnover of the rotation curve occurs at only $\approx$55\,pc from the nucleus, suggesting that the stellar motions are dominated by a highly concentrated gravitational potential. This is also supported by the small value obtained for the scale length ($A=39.7$\,pc) from the modelling of the velocity field. The derived parameters are in approximate agreement with the ones derived by \citet{barbosa06} from a similar modelling using optical IFU data for a field-of-view of $\approx7^{\prime\prime}\times$5$^{\prime\prime}$ (315$\times$225\,pc$^2$). The exception is our value of $\Psi_0=120^\circ$, which is $\sim13^\circ$ larger than theirs. On the other hand, our $\Psi_0$ is $\sim13^\circ$ smaller than the one derived by \citet{dumas07} from stellar kinematics obtained over the much larger field-of-view of $\approx50^{\prime\prime}\times$40$^{\prime\prime}$ (2250$\times$1800\,pc$^2$). We atribute these differences to the differences in field-of-view, considering also that our field-of-view, which corresponds to  130$\times$180\,pc$^2$ at the galaxy, is only sampling the stellar kinematics very close to the nucleus. 

For an orientation of the line of nodes of $\Psi_0=120^\circ$, and assuming that the spiral arms of NGC\,4051 are trailing (orientation of the arms is shown in Fig.\,\ref{espectro}), we conclude that the NE is the far side and the SW is the near side of the galaxy. 

The stellar velocity dispersion values show patches with lower values ($\approx$40-50\,km\,s$^{-1}$) on top of a common background of values ranging from 60 to 70\,km\,s$^{-1}$. A possible interpretation is that these patches are colder regions with more recent star formation than the underlying bulge. This is supported by optical spectra of the nuclear region which show clear signatures of intermediate age stars \citep{cid-fernandes03}.
A very small $\sigma$ drop is observed right at the nucleus, but we take this result with caution because of the obvious contamination of the spectra of the nuclear region by emission from dust and broad emission lines.

We can estimate the mass of the SMBH ($M_{BH}$) from the bulge stellar velocity dispersion ($\sigma_*$) as ${\rm log}(M_{BH}/{\rm M_\odot})=\alpha+\beta\,{\rm log}(\sigma_*/\sigma_0)$, where $\alpha=8.13\pm0.06$, $\beta=4.02\pm0.32$ and $\sigma_0=200$\,km\,s$^{-1}$ \citep{tremaine02}. Adopting $\sigma_*\approx60$\,km\,s$^{-1}$ as representative of the bulge (top right panel of Fig.\,\ref{stel}), we obtain $M_{BH}=1.1\pm0.3\times10^6$\,M$_\odot$.  This value is in good agreement with those obtained by previous authors from reverberation mapping and scaling relations \citep{shemmer03,kaspi00}. For this mass, the radius of influence of the SMBH is $\approx1.3\,$pc, thus not resolved at the spatial resolution of our data.

\subsection{Gas Kinematics} \label{kinematics}

The simultaneous observation of the stellar and gaseous kinematics allowed us to construct  residual maps for the gaseous kinematics relative to the stellar kinematics model described in section\,\ref{secplum}. The residual map for the H$_2$ emitting gas is presented in Fig.\,\ref{res}. The most conspicous feature in this map is the elongated structure to the NE, which shows  blueshifts of up to  $\approx\,-$100\,km\,s$^{-1}$, also seen in the corresponding velocity slices in Fig.\,\ref{slices}. These blueshifts could be either due to an inflow, if the gas is located in the plane, as the NE is the far side of the galaxy; or to an outflow, if the gas is extended to high latitudes (above the plane), e.g. in a conical structure oriented towards us. We discuss below these two possibilities.

The outflow interpretation is supported by the following facts: (1) that the elongated NE structure is {\it approximately} oriented along the radio jet; (2)  a {\it similar} structure and blueshifts have been observed by \citet[][-hereafter B08]{barbosa07} in IFU observations of [S\,{\sc iii}]$\,\lambda9069$  emitting gas, who found, in addition, redshifts to the SW, which appear to be due to a counterpart conical outflow related to the SW part of the radio-jet, probably located behind the galactic plane and being directed away from us; (3) the velocity dispersion and line ratio maps which show increased values in regions close to the tips of the radio jet.

The inflow interpretation is, on the other hand, supported by: (1) the fact that the orientation of the NE blueshifted region is not well aligned with the direction of the radio jet, but is shifted by $\sim$0\farcs5 to the S--SE (see Fig.\,\ref{slices}); (2) the fact that the blueshifted region is elongated and curved as  if it belonged to a spiral arm ending at the nucleus, while the blueshifted structure observed in the [S\,{\sc iii}]$\,\lambda9069$  emitting gas by B08 shows a more conical shape whose axis is better aligned with the direction of the radio jet; (3) the fact that there is no redshifted counterpart in H$_2$ as observed for the [S\,{\sc iii}] emitting gas. There is instead a redshifted structure observed in the velocity slices (see Fig.\,\ref{slices}) also curved as belonging to a spiral arm in the near side (SW) of the galaxy. 

Considering the arguments above we favour the inflow interpretation. The blueshifts to the NE and redshifts to the SW observed in the H$_2$ emitting gas would then be due to inflow of molecular gas in the galaxy plane along nuclear spiral arms. The presence of such arms is supported by the structure map of the nuclear region of NGC\,4051 presented by \citet{lopes07}, which shows similar spiral structure with the same curvature and orientation of the arms in an HST optical image of the nuclear region of NGC\,4051. The structure map also shows a lot of obscuration against the near side of the galaxy, which would explain why the flux of the redshifted gas emission is fainter than  that of the blueshifted gas on the far side of the galaxy.
The enhancements in the gas velocity dispersions and increase in line emission ratios observed in H$_2$ and Br$\gamma$ emitting gas can be understood if the radio jet is launched at a relatively small angle to the galaxy plane compressing the circumnuclear ISM close to the nucleus. 



In a recent  study aimed at investigating the interaction between the radio jet and line emission in the NLR of the Seyfert galaxy ESO\,428-G14 using GNIRS IFU data we found a close association of the [Fe\,{\sc ii}] and H\,{\sc i}  emission line distributions and  kinematics with the radio emission distribution \citep{riffel06}, a result also found in some studies  \citep[eg.][B08]{falcke98,bicknell00,tilak05}, but also contested by others  \citep[eg.][]{kaiser00,das05}. In ESO\,428-G14,  \citet{riffel06} concluded that the H$_2$ kinematics was distinct from that in the other emission-lines: although part of the H$_2$ emitting gas was affected by interaction with the radio jet, most of it was located in the galaxy plane and was not affected by the radio jet. A less disturbed kinematics for H$_2$ than for other emission lines was also found in a near-IR study of the gas emission around the nuclei of the Seyfert galaxies NGC\,2110 and Circicus \citep{storchi-bergmann99}. These studies thus also support our favored interpretation in the present study: most of the H$_2$ emission is located in the plane and is not related to the outflow. It is instead, flowing towards the nucleus along spiral arms.

The Br$\gamma$ emitting gas is almost all at the systemic velocity as observed in the bottom-left panel of Fig.\,\ref{gas} and seems to show no rotation, which leads us to conclude that most of the Br$\gamma$ emitting gas is  not restricted to the plane of the galaxy.

\subsection{[Ca\,{\sc viii}] coronal emission}

Coronal lines are forbidden transitions from highly ionized species which extend from the unresolved nucleus up to distances between a few tens to a few hundreds of parsecs, and usually present blue wings and are broader than low ionization lines \citep[e.g.][and references therein]{rodriguez-ardila06}. Thus, it is possible to resolve the coronal line region for the closest active galaxies. In NGC\,4051, the [Ca\,{\sc viii}] coronal emission line at 2.3211\,$\mu$m is indeed resolved, presenting extended emission in a circular region around the nucleus with diameter of $\approx$75\,pc. Blueshifts of up to  $-250\,$km\,s$^{-1}$ are observed very close to the nucleus (within the inner $\approx$25\,pc), a value that is much higher than that observed for Br$\gamma$ and  H$_2$ emission lines. The mean velocity dispersion is 150\,km\,s$^{-1}$, which is about 2--3 times higher than that obtained  the other lines as well.  This kinematics together with the compact flux distribution supports the interpretation that the coronal lines are produced close to the nucleus, and probably in the transition region between the BLR (broad-line region)  and the NLR as suggested by previous authors. The spatial extent of the emission, the blueshifts and the $\sigma$ values observed in NGC\,4051 are similar to the ones obtained for the Seyfert\,2 galaxy Circinus, while for more active galaxies such values tend to be higher \citep{rodriguez-ardila06}.

\subsection{Mass of the emitting gas}

\begin{figure}
\centering
\includegraphics[scale=1.35]{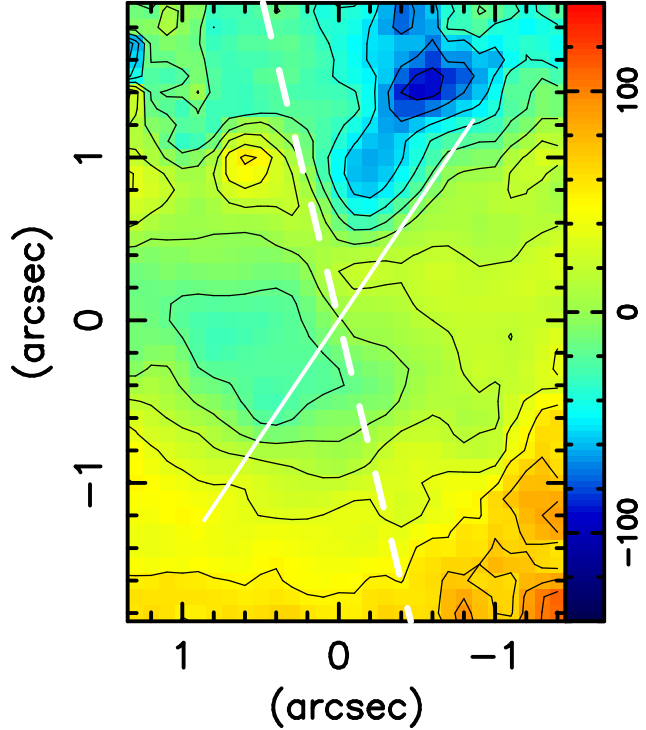}
\caption{Difference between the H$_2$ and stellar kinematics. The white dashed line shows the orientation of the line of nodes and the full white line shows the PA of the radio emission.} 
\label{res}
\end{figure}

The mass of hot H$_2$ can be estimated as \citep[e.g.][]{scoville82}:
\[
 M_{H_2}=\frac{2m_p\,F_{H_{2}\lambda2.1218}\,4\pi d^2}{f_{\nu=1,J=3}A_{S(1)}\,h\nu}
\]
\begin{equation}
~~~~~~~=5.0776\times10^{13}\left(\frac{F_{H_{2}\lambda2.1218}}{\rm erg\,s^{-1}\,cm^{-2}}\right)\left(\frac{d}{\rm Mpc}\right)^2,
\label{mh2}
\end{equation}
where $m_p$ is the proton mass, $F_{H_2\lambda2.1218}$ is the line flux, $d$ is the galaxy distance and $M_{H_2}$ is given in solar masses. As proposed by \citet{scoville82}, we assume a typical vibrational temperature of $T_{vib}=2000$\,K, which implies a population fraction $f_{\nu=1,J=3}=1.22\times10^{-2}$ and a transition probability $A_{S(1)}=3.47\times10^{-7}$\,s$^{-1}$ \citep{turner77}. The line flux should be corrected for the intrinsic $E(B-V)$, but for NGC\,4051, $E(B-V)$ is small \citep{rogerio06} and thus we do not apply any correction.  For the E--NE inflow region we obtain $F_{H_2\lambda2.1218}=2.1\times10^{-15}$\,erg\,s$^{-1}$\,cm$^{-2}$ integrated within the approximately triangular region which includes blueshifts larger than $-100$\,km\,s$^{-1}$ in Fig.\,\ref{res} and obtain a mass of $M_{H_2} \approx {\rm 9.3\,M_\odot}$. 
Integrating now the H$_2\,\lambda2.1218\,\mu$m emission over the same field where the Br$\gamma$ is observed (the central 130$\times$135\,pc$^2$) we obtain $F_{H_2\lambda2.1218}=1.5\times10^{-14}$\,erg\,s$^{-1}$\,cm$^{-2}$ and thus $M_{H_2}\approx66\,{\rm M_\odot}$. 

The above masses are small, but we point out that in the nuclear region of galaxies it has been shown that the hot-to-cold mass ratio ranges between 10$^{-7}$ to 10$^{-5}$ \citep{dale05}, suggesting that the total mass of molecular gas can be orders of magnitude larger than the value measured directly from the H$_2$ emission.

The mass of the ionized hydrogen can be estimated as $M_{HII}=m_p\,N_e\,V_{HII}$, where $N_e$ is the electron density and $V_{HII}$ is the volume of the emitting region. Replacing $N_e\,V_{HII}$ by the expression given in \citet{scoville82} we obtain:
\begin{equation}
 M_{HII}=2.88\times10^{17}\left(\frac{F_{\rm Br\gamma}}{\rm erg\,s^{-1}cm^{-2}}\right)\left(\frac{d}{\rm Mpc}\right)^2,
\end{equation}
where $M_{HII}$ is given is solar masses, and we assume an electron temperature $T=10^4$\,K and electron density $N_e=100\,{\rm cm^{-3}}$. The total integrated Br$\gamma$ flux is  $F_{\rm Br\gamma}\approx5.8\times10^{-15}\,{\rm erg\,s^{-1}\,cm^{-2}}$ and thus $M_{HII}\approx1.4\times10^5\,{\rm M_\odot}$. The mass of ionized gas is thus about 2000 larger than the mass of hot H$_2$ gas.

 

Using the above calculated mass $M_{H_2}$ for the E--NE blueshifted region we can estimate the flux of inflowing hot molecular gas under the assumption that this gas is streaming towards the nucleus. We calculate the hot ${H_2}$ mass crossing a circular cross section by $\dot{M}_{H_2}=2m_p\,N_{H_2}\,v\,\pi r^2$, where $v$ is the inflowing velocity and $r$ is the radius of the circular cross section. The maximum blueshift velocities occur at 1\farcs2 E-NE of the nucleus where the cross-section radius of the structure shown in Fig.\,\ref{res} is estimated to be $r=$0\farcs55$\,\approx25$\,pc. Assuming $v=100$\,km\,s$^{-1}$ and conical geometry to calculate the H$_2$ density, we obtain $\dot{M}_{H_2}=2.5\times10^{21}$\,g\,s$^{-1}\approx4\times10^{-5}{\rm \, M_\odot\,yr^{-1}}$. This rate is probably up to 2 times larger if we consider a similar inflow along the W-SW spiral arm, observed in redshift.

The mass accretion rate necessary to power the active nucleus can be estimeted by $\dot{M}=\frac{L_{\rm bol}}{c^2\eta}$, where $L_{\rm bol}$ is the bolometric luminosity, $c$ is the light speed and $\eta$ is the efficiency of conversion of the rest mass energy of the accreted material into radiation. Assuming a typical value of $\eta\approx0.1$ and using $L_{\rm bol}=2.7\times10^{43}\,{\rm erg s^{-1}}$ from \citet{ogle04} we obtain $\dot{M}=4.7\times10^{-3}\,{\rm M_\odot/yr}$. The H$_2$ mass inflow rate calculated above  is then about 100 times smaller than the accretion rate needed to power the AGN of NGC\,4051, supporting additional contribution of cold non-emitting gas to the mass inflow rate.

\subsection{H$_2$ excitation}


The H$_2$ lines can be excited by three distinct processes: fluorescent
excitation through absorption of soft-UV photons (912--1108 \AA) in
the Lyman and Werner bands \citep{black87},  excitation by X-ray heating \citep{maloney96} and excitation by shocks  \citep{hollembach89}. The first is usually considered a non-thermal process while the other two are considered  thermal processes.  Several studies investigated the H$_2$ excitation mechanisms \citep[ eg.][]{draine90,veilleux97,quillen99,bellamy04,rodriguez-ardila05,rogerio06,riffel06,zuther07}.


\citet{quillen99} have looked for correlations  of the H$_2$ emission with radio 6\,cm and
hard X-ray fluxes. They found no correlation with X-rays, suggesting
X-rays heating is not the dominant H$_2$ excitation mechanism, and found a weak
correlation with radio 6\,cm, suggesting that no single mechanism is
likely to be responsible for the molecular hydrogen excitation in
Seyfert galaxies.  \citet{draine90} propose that most of the H$_2$ line emission originates in molecular gas which is heated by transient X-ray irradiation. Hard X-rays from the AGN also have been  proposed by \citet{wilman00} and \citet{bellamy04} as the dominant excitation mechanism of H$_2$ emission in Cygnus\,A.  \citet{zuther07} using the \citet{maloney96} X-rays excitation models found that the X-ray emission can account for  some H$_2$ excitation for the starburst/Seyfert galaxy Mrk\,609.  For the case of ESO428-G14 we concluded that shocks produced by a radio jet play a fundamental role in the kinematics and excitation of the H$_2\lambda$2.1218 emission line in regions co-spatial with the radio emission and X-rays are important in regions away from the radio structures \citep{riffel06}.

The H$_2\,\lambda2.2477/\lambda2.1218$ line ratio is commonly used to distinguish between thermal ($\sim$0.1-0.2) and fluorescent ($\sim$0.55) excitation mechanisms for the H$_2$ emitting gas \citep{mouri94,reunanen02,rodriguez-ardila05}, while the H$_2\,\lambda2.0338/\lambda2.2235$ ratio is used to determine the thermal excitation temperature. The H$_2\,\lambda2.2477\,\mu$m emission line is only marginally detected in our individual spectra, so we integrated the flux of this line over the whole IFU field obtaining H$_2\,\lambda2.2477/\lambda2.1218 \approx0.12\pm0.02$ indicating that the emitted H$_2$ is excited by thermal processes.

The rotational and vibrational temperatures for the molecular gas can also be used to distinguish between thermal and fluorescent excitation: for thermal excitation the vibrational and rotational temperatures must be similar and for fluorescent excitation the vibrational temperature must be higher  than the rotational temperature -- non-local UV photons overpopulate the highest energy levels compared to the population distribution  expected for a Maxwell-Boltzmann distribution. The rotational temperature can be obtained by 
$T_{\rm rot} \cong -{\rm 1113/ln(0.323\,}\frac{F_{H_{2}\lambda2.0338}}{F_{H_{2}\lambda2.2235}})$ 
and the vibrational temperature by 
$T_{\rm vib}\cong {\rm 5600/ ln(1.355\,}\frac{F_{H_{2}\lambda2.1218}}{F_{H_{2}\lambda2.2477}})$  \citep{reunanen02}.
For NGC\,4051 we obtain H$_2\,\lambda2.0338/\lambda2.2235=1.8\pm0.28$ and H$_2\,\lambda2.1218/\lambda2.2477=8.4\pm1.02$, thus $T_{\rm rot} = 2052^{+746}_{-488}$\,K and $T_{\rm vib} = 2305_{-120}^{+157}$\,K. The similarity of these two values reinforce thermal processes as the dominant excitation mechanism of the molecular hydrogen. The value obtained for the vibrational temperature supports the assumption ($T_{\rm vib}=2000\,{\rm K}$) used in equation\,\ref{mh2} to obtain the hot H$_2$ mass.


In order to test if X-ray emission can account for the excitation of the H$_2\,\lambda$2.1218$\,\mu$m emission line for NGC\,4051 we have used the models of \citet{maloney96} to estimate the emergent H$_2$ flux of a gas cloud illuminated by a source of hard X-rays with an intrinsic luminosity $L_X$. The cooling is given by the effective ionization parameter $\xi_{\rm eff}$ \citep{zuther07}:
\begin{equation}
\xi_{\rm eff}={\rm 1.26\times10^{-4}}\frac{f_X}{n_5N_{22}^{0.9}},
\end{equation}
where $f_X$ is the incident hard X-ray flux at the distance $d$[pc] from the X-ray source, $n_5$[10$^{-5}$cm$^{-3}$] is the total hydrogen gas density, and $N_{22}$[10$^{22}$cm$^2$]  is the attenuating column density.  Using  $f_X$=2.3$\times$10$^{-11}$\,erg\,cm$^{-2}$\,s$^{-1}$ and $N_H$=1.32$\times$10$^{20}$\,cm$^{-2}$ obtained from the ASCA Tartarus database we can calculate  $\xi_{\rm eff}$.  \citet{maloney96} calculate emergent fluxes for two gas densities, 10$^5$\,cm$^{-3}$ and 10$^3$\,cm$^{-3}$. We calculate the effective ionization parameter for three different distances of the AGN (25, 50 and 75\,pc), and for the same gas densities of \citet{maloney96}. Using this and their Fig.\,6(a,b) we can obtain the emergent intensity of the H$_2\,\lambda$2.1218$\,\mu$m emission line. In Table\,\ref{fluxes}  we present the calculated effective ionization parameter and the emergent flux for an aperture of 0\farcs1$\times$0\farcs1 -- corresponding to a solid angle 2.34$\times$10$^{-13}$\,sr -- in erg\,cm$^{-2}$\,s$^{-1}$. In this table we also present the observed H$_2\,\lambda2.1218\,\mu$m obtained by the average of the H$_2$ flux in a ring with radius $d$ and width of 0\farcs2.

\begin{table}
\caption{Comparasion of the observed H$_2\,\lambda2.1218\,\mu$m fluxes and calculated using models of \citet{maloney96} for an aperture of 0\farcs1$\times$0\farcs1 for hydrogen densities $n=10^5$\,cm$^{-3}$ and $n=10^3$\,cm$^{-3}$ .}
\centering
\begin{tabular}{c c c c c c}
\hline
  & Observed & \multicolumn{2}{c}{$\rm n=10^5\,cm^{-3}$}&   \multicolumn{2}{c}{$\rm n=10^3\,cm^{-3}$}     \\
\hline
$d$  & \small{log($F_{H_2}$) }&   log($\xi_{\rm eff})$ & log($F_{H_2}$)   &   log($\xi_{\rm eff})$  & log($F_{H_2}$)  \\
25    &   -16.4 &    -1.7            & -14.7           &              --                 &           --        \\
50    &   -17.0 &    -2.3            & -17.2           &             -0.3                 &           -15.8         \\
75    &   -17.3 &    -2.7            & -17.0           &             -0.7                 &          -15.7          \\
\hline
\end{tabular}
\label{fluxes}
\end{table}

By comparing the emergent fluxes from Table\,\ref{fluxes} obtained by X-ray heating models with the observed fluxes we conclude that excitation by X-ray heating can account for most of the observed H$_2$ flux. 

Another line ratio commonly used to investigate the H$_2$ excitation is H$_2\,\lambda2.1218$/Br$\gamma$. In starburst galaxies, where the main heating agent is UV radiation, H$_2$/Br$\gamma<0.6$ \citep{rodriguez-ardila05}, while for Seyferts this ratio is larger ($0.6<$H$_2$/Br$\gamma<2$) because of additional H$_2$ emission excited by shocks or by X-rays from the AGN \citep{rodriguez-ardila05}. As observed in the bottom-right panel of Fig.\,\ref{flux}, NGC\,4051 presents H$_2$/Br$\gamma=1\pm 0.2$ at the nucleus, which is a typical value for AGNs. The highest values reach H$_2$/Br$\gamma=8\pm 1.9$ and are observed E-NE and to the W-SW of the nucleus, close to the tips of the radio jet, supporting some contribution of shocks by the radio jet to the excitation of the molecular hydrogen in these regions. On the other hand, the interpretation of this line ratio should be considered with caution due to the fact that the bulk of H$_2$ emission seems to originate from the disk while the Br$\gamma$ emission seems not to be restricted to the disk.



\section{Conclusions}

We have analysed two-dimensional near-IR $K-$band spectra from the inner $\approx$\,150\,pc of the Narrow Line Seyfert 1 galaxy NGC\,4051 obtained with the Gemini NIFS integral field spectrograph at a sampling of 4.5$\times$4.5\,pc$^2$ at the galaxy and spectral resolution of $\approx$3\,\AA. We have mapped the stellar and gaseous kinematics, and the emission-line flux distributions and ratios of the molecular and ionized hydrogen. The main conclusions of this work are:

\begin{itemize}

\item The turnover of the stellar rotation curve is at only $\approx$55\,pc from the nucleus, suggesting that the stellar motions are dominated by a highly concentrated gravitational potential, a result confirmed by modeling using a Plummer gravitational potential, as we obtain a small scale length, $A\approx$39\,pc. This result supports the findings of \citet{barbosa06} based of optical data. The mean velocity dispersion of the bulge ($\sim$60\,km\,s$^{-1}$) implies a SMBH mass of $\sim 10^6$\,M$_\odot$. Within the bulge, we find patches of lower velocity dispersions, which we attribute to recent star formation.

\item The gas kinematics is distinct for each emission line: the Br$\gamma$ emission line shows velocities restricted to within $\sim$30\,km\,s$^{-1}$ from systemic, without evidence of bultk motions, suggesting that the ionized gas is not restricted to the galactic plane, while  much larger blueshifts and redshifts are observed for both the [Ca\,{\sc viii}] and H$_2$ emitting gas. The Br$\gamma$ velocity dispersion is small ($\sim$40\,km\,s$^{-1}$) over most of the field, except around the nucleus at a location close to the SW tip of the radio jet, suggesting some energy injection in the gas by interaction with the radio jet. 

\item The [Ca\,{\sc viii}] coronal emission line is compact but resolved, extending over a circular region of radius  $\approx$35\,pc around the nucleus. It also present the largest velocity blueshifts ($-250\,$km\,s$^{-1}$) and velocity dispersion (150\,km\,s$^{-1}$) of the emission lines studied here, supporting an origin close to the active nucleus, possibly in the transition region between the BLR and the NLR.

\item The H$_2$ emitting gas seems to be mostly restricted to the galaxy plane. The most conspicuous kinematic features are a curved elongated blueshifted structure to the NE, interpreted as gas inflow along a nuclear spiral arm in the far side of the galaxy, and a curved, redshifted structure to the SW, interpreted as gas inflow along a nuclear spiral arm in the near side of the galaxy. Estimates of the mass inflow rate in hot H$_2$ gives $\dot{M}_{H_2} \approx 4\times10^{-5}\,{\rm M_\odot\,yr^{-1}}$, which is $\sim\,100$ times smaller than the nuclear accretion rate necessary to power the active nucleus of NGC\,4051, supporting the presence of additional inflow of cold non-emitting gas.  This is not the first time we have been able to map inflows towards an active nucleus along nuclear spiral arms. In two previous studies we \citep{fathi06,storchi-bergmann07} have mapped streaming motions towards the active nucleus along nuclear spirals in the galaxies NGC\,1097 and NGC\,6951 although in ionized gas. As nuclear spirals are ubiquitous around active galactic nuclei \citep{lopes07,prieto05}, such inflows seem to be an ``universal'' mechanism to bring gas inwards to feed the SMBH within the inner few hundred parsecs of the galaxy.


\item The total mass of hot H$_2$ is estimated to be of the order of 66\,M$_\odot$, while that of H\,{\sc ii} is estimated to be 1.4$\times\,10^5$ M$_\odot$.

\item From the H$_2$ line ratios we conclude that H$_2$ is excited by thermal processes -- heating by X-rays from the AGN and shocks produced by the radio jet. We conclude, based in X-ray excitation models of \citet{maloney96} that X-ray heating can account for the observed emission, but the H$_2\,\lambda2.1218\,\mu$m/Br$\gamma$ line ratio supports some contribution from shocks in the regions where the radio jet interacts with the H$_2$ emitting gas.

\end{itemize}

\section*{Acknowledgments}
Based on observations obtained at the Gemini Observatory, which is operated by the Association of Universities for Research in Astronomy, Inc., under a cooperative agreement with the NSF on behalf of the Gemini partnership: the National Science Foundation (United States), the Science and Technology Facilities Council (United Kingdom), the National Research Council (Canada), CONICYT (Chile), the Australian Research Council
(Australia), CNPq (Brazil) and SECYT (Argentina). 
Basic research in astronomy at the NRL is supported by NRL 6.1 Base funding.
The National Radio Astronomy Observatory is a facility of the National Science Foundation operated under cooperative agreement by Associated Universities, Inc.
This research has made use of the NASA/IPAC Extragalactic Database (NED) which is operated by the Jet Propulsion Laboratory, California Institute of Technology, under contract with the National Aeronautics and Space Administration.
This research has made use of the Tartarus database, created by Paul O'Neill and Kirpal Nandra at Imperial College London, and Jane Turner at NASA/GSFC. Tartarus is supported by funding from PPARC, and NASA grants NAG5-7385 and NAG5-7067

\label{lastpage}

\end{document}